\documentclass[%
 reprint,
 superscriptaddress,
 amsmath,amssymb,
 aps,
 pra,
]{revtex4-2}

\usepackage{graphicx}% Include figure files
\usepackage{dcolumn}% Align table columns on decimal point
\usepackage{bm}% bold math
\usepackage[hidelinks]{hyperref}% add hypertext capabilities
\usepackage{siunitx}
\usepackage{xspace}
\newcommand{\Yb}{\textsuperscript{171}Yb\textsuperscript{+}\xspace}
\newcommand{\Sr}{\textsuperscript{88}Sr\textsuperscript{+}\xspace}

\newcommand\T{\rule{0pt}{2.6ex}}       % Top strut
\newcommand\B{\rule[-1.2ex]{0pt}{0pt}} % Bottom strut

\begin{document}

\preprint{APS/123-QED}

\title{A multi-ion optical clock with $\mathbf{5 \times 10^{-19}}$ uncertainty}

\author{Melina Filzinger}%\email{iauthor@gmail.com}
\affiliation{Physikalisch-Technische Bundesanstalt, Bundesallee 100, 38116 Braunschweig, Germany}

\author{Martin R. Steinel}%\email{iiauthor@gmail.com}
\affiliation{Physikalisch-Technische Bundesanstalt, Bundesallee 100, 38116 Braunschweig, Germany}

\author{Jian Jiang}
\affiliation{Physikalisch-Technische Bundesanstalt, Bundesallee 100, 38116 Braunschweig, Germany}

\author{Daniel Bennett}
\affiliation{Physikalisch-Technische Bundesanstalt, Bundesallee 100, 38116 Braunschweig, Germany}

\author{Tanja E. Mehlstäubler}
\affiliation{Physikalisch-Technische Bundesanstalt, Bundesallee 100, 38116 Braunschweig, Germany}
\affiliation{Institut für Quantenoptik, Leibniz Universität Hannover, Welfengarten~1, 30167 Hannover, Germany}

\author{Ekkehard Peik}
\affiliation{Physikalisch-Technische Bundesanstalt, Bundesallee 100, 38116 Braunschweig, Germany}

\author{Nils Huntemann}\email{nils.huntemann@ptb.de}
\affiliation{Physikalisch-Technische Bundesanstalt, Bundesallee 100, 38116 Braunschweig, Germany}

\date{\today}% It is always \today, today,

\begin{abstract}
Today's most accurate clocks are based on laser spectroscopy of electronic transitions in single trapped ions and feature fractional frequency uncertainties below $1\times10^{-18}$. Scaling these systems to multiple, simultaneously interrogated ions reduces measurement times, driving recent advances in multi-ion clocks. However, maintaining state-of-the-art systematic uncertainties while increasing the number of ions remains a central challenge. Here, we report on a multi-ion optical atomic clock with a fractional frequency uncertainty of $5.3\times10^{-19}$ and up to 10 \Sr ions. Ion-resolved state detection enables minimization of position-dependent shifts, with residual effects suppressed below the $10^{-20}$-level. Clock operation with eight to ten ions reduces the measurement time by a factor of 4.8 compared to single-ion operation. A comparison with an established \Yb single-ion clock yields an unperturbed frequency ratio of $0.6926711632159660405(20)$, with a statistical uncertainty of $0.9\times10^{-18}$ and a combined uncertainty of $2.9\times 10^{-18}$. These results demonstrate robust multi-ion clock operation with reduced averaging time and state-of-the-art accuracy.
\end{abstract}

\maketitle

%\tableofcontents

Optical atomic clocks based on neutral atoms in optical lattices or ions in radiofrequency traps are the most precise measurement instruments developed to date. Their performance is now surpassing that of the current best primary frequency standards by more than two orders of magnitude, motivating an anticipated optical redefinition of the second~\cite{Dimarcq2024}. Besides their applications as frequency standards, optical clocks are powerful tools for tests of fundamental physics~\cite{Safronova2018}. Single-ion optical clocks have recently reported fractional systematic uncertainties in the $10^{-19}$-range~\cite{Marshall2025, Zhang2025, Lindvall2025b, Zhang2023}.
However, their frequency instability is typically limited by the large quantum projection noise intrinsic to measurements with a single ion, making long averaging times necessary for small statistical uncertainties. 
Simultaneously interrogating multiple ions in a linear Coulomb crystal offers a conceptually simple avenue towards an improved frequency instability and therefore reduced measurement times~\cite{Herschbach2012}. Here, inhomogeneities in frequency shifts across such an ion chain need to be considered and controlled~\cite{Keller2019, Keller2019a}. So far, clock operation with three \Sr ions~\cite{Steinel2023}, seven \Sr ions~\cite{Akerman2025}, and four \textsuperscript{115}In\textsuperscript{+} ions~\cite{Hausser2025} has been demonstrated, though not with fully evaluated systematic uncertainties.

Here, we report on a multi-ion optical clock, based on the ${}^2S_{1/2}\leftrightarrow {}^2D_{5/2}$ transition in \Sr, with a systematic uncertainty of $5.3\times10^{-19}$. The employed transition is well-established as the basis for optical clocks~\cite{Dube2013, Steinel2023, Lindvall2025b}, featuring a convenient wavelength, a well-characterized sensitivity to thermal radiation~\cite{Lindvall2025a}, and the possibility to suppress motional shifts~\cite{Dube2014}. However, its first-order magnetic sensitivity and large quadrupole moment~\cite{Barwood2004} have made multi-ion implementations challenging. The resulting inhomogeneous frequency shifts have previously been suppressed to the $10^{-17}$-level using dynamical decoupling~\cite{Shaniv2019, Akerman2025}. Here, we instead rely on static suppression and show that residual effects of frequency shift variations along a chain of up to ten ions are below the $10^{-20}$ level. In fact, multi-ion operation reduces the systematic uncertainty from $6.5\times10^{-19}$ for a single ion due to a reduced uncertainty from background gas collisions. We compared the new optical clock to a single-ion clock based on the $^2S_{1/2}\leftrightarrow{}^2F_{7/2}$ transition in \Yb~\cite{Huntemann2016, Sanner2019}, obtaining a combined fractional uncertainty of $2.9\times10^{-18}$, dominated by the \Yb clock, on the resulting frequency ratio.

\begin{figure*}
\centering
\includegraphics[width=\textwidth]{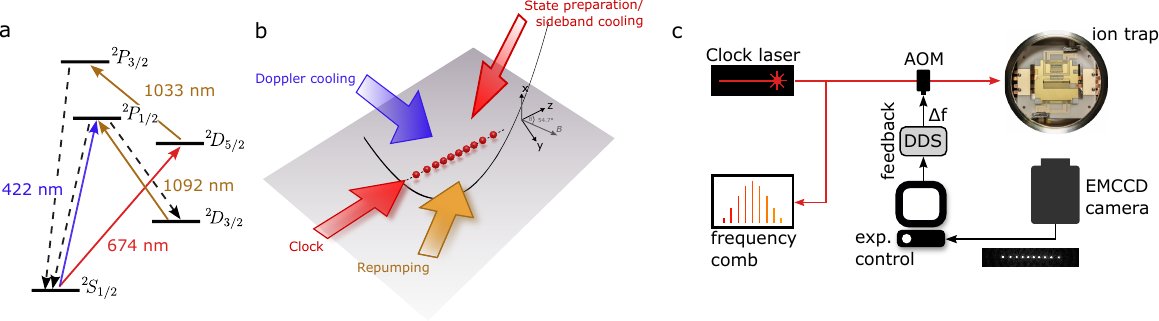}
\caption{Overview over experiment. a) Relevant level scheme of \Sr and b) overview over measurement geometry: A chain of ions is trapped in a linear radiofrequency trap. The ions are addressed collectively: laser beams with elliptical cross-section are used for Doppler cooling and state detection (422\,nm), state preparation and sideband cooling (674\,nm), and repumping (1033\,nm and 1092\,nm). For clock operation, the ions are interrogated on the ${}^2S_{1/2}\leftrightarrow {}^2D_{5/2}$ clock transition with a 674\,nm laser beam propagating along the trap axis. The magnetic field defining the quantization axis is set with an angle of 54.7° w.r.t. the trap axis to suppress the effect of varying electric field gradients along the ion chain (see text for details). c) Key components of the multi-ion optical clock: A narrow-linewidth clock laser is used to repeatedly interrogate the trapped ions on the clock transition. Successful excitation is detected in an ion-resolved fashion using an electron-multiplying charge-coupled device (EMCCD) camera. The experimental control sequence steers the global frequency offset $\Delta f$, which is applied using an acousto-optic modulator (AOM), via direct digital synthesis (DDS). A frequency comb, together with the time-resolved recording of $\Delta f$, enables frequency comparisons with other optical clocks.}\label{fig1}
\end{figure*}

A single \Sr ion or a linear Coulomb crystal of eight to ten \Sr ions is confined in one segment of a linear radiofrequency (RF) trap based on the design in \cite{Keller2019}. An overview over the experimental configuration is shown in \autoref{fig1}. 
Elliptical laser beams for cooling, repumping, and state preparation ensure uniform addressing of all ions in the chain. A dedicated high-intensity beam at 674\,nm drives the clock transition for fast state preparation and sideband cooling. A weaker clock laser beam is aligned along the ion chain, yielding the same Rabi frequency for all ions. 

During clock operation, a local frequency offset to a narrow-linewidth clock laser is steered based on repeated interrogation of the reference transition. For this, the same experimental sequence is repeated periodically: Laser cooling on the dipole-allowed $^2S_{1/2} \leftrightarrow\, ^2P_{1/2}$ transition at 422\, nm for 5\,ms reduces the ions' thermal motion close to the Doppler limit, with laser light at $1092$\,nm preventing population trapping in the $^2D_{3/2}$ state. For multiple ions, we additionally perform resolved sideband cooling of the axial centre-of-mass (COM) motional mode. Subsequent preparation in the $^2S_{1/2}, m_J=-1/2\, (+1/2)$ sublevel is achieved with multiple short pulses on the $^2S_{1/2}, m_J=+1/2\,(-1/2)\rightarrow \,^2D_{5/2}, m_J=- 3/2\, (+3/2)$ transition respectively, with repumping from the $\,^2D_{5/2}$ state after each pulse~\cite{Steinel2023}. After mechanical shutters block the light used for cooling and preparation, Rabi interrogation of the clock transition is performed. A successful excitation is indicated by absence of fluorescence during subsequent illumination with cooling laser light. After detection, laser light at 1033\,nm enables fast depletion of the excited clock state via the $^2P_{3/2}$ state back to the ground state~\cite{Dube2015}. A clock interrogation cycle is valid (used for steering of the clock laser) if the fluorescence signal reappears within the subsequent cooling period. An invalid cycle is repeated, and lack of fluorescence over several cycles triggers a re-cooling sequence that involves reducing the rf and dc trap confinement and red-detuning the 422\,nm laser as it is illuminating the ion(s). While we use a photomultiplier tube for state detection of a single ion, we use a camera for detection during multi-ion operation. Image processing and data transfer add up to 15\,ms of dead time to the clock sequence after Doppler cooling. 

The $^2S_{1/2}\leftrightarrow \,^2D_{5/2}$ transition is split into ten first-order Zeeman-sensitive components via a static external magnetic field of about $3\,\text{\textmu T}$. We interrogate opposite-sensitivity pairs of these Zeeman components, and use the resulting average, which is free of the first-order Zeeman shift. Additionally,  a weighted average of two such transition pairs suppresses tensorial frequency shifts~\cite{Dube2013}. Magnetic field noise limits the maximum coherent interrogation time in our setup, and consequently we scale the interrogation time inversely with the Zeeman sensitivity of each transition pair: 
$250\,$ms long interrogation pulses are used on each of the $^2S_{1/2}, m_J=\pm 1/2\leftrightarrow \,^2D_{5/2}, m_J=\pm 3/2$ transitions, and $100\,$ms pulses on the $^2S_{1/2}, m_J=\pm 1/2\leftrightarrow \,^2D_{5/2}, m_J=\pm 5/2$ transitions. We observe Fourier-limited linewidths and maximum excitation probabilities of about 70\% and 80\% respectively, with the difference resulting from the 391\,ms~\cite{Letchumanan2005} excited-state lifetime.
We use the sequence we have previously found to yield the best frequency instability of a \Sr clock limited by magnetic field noise~\cite{Steinel2023}: The $^2S_{1/2}, m_J=\pm 1/2\leftrightarrow \,^2D_{5/2}, m_J=\pm 3/2$ transition pair is interrogated twice as often as the $^2S_{1/2}, m_J=\pm 1/2\leftrightarrow \,^2D_{5/2}, m_J=\pm 5/2$ transition pair, and they are averaged with respective weights of 5/6 and 1/6. 

We confine chains of ions oriented along the RF-null (axial direction) of the linear ion trap. When operating the clock with a single ion, we typically employ secular trapping frequencies of about $\omega_\text{rad1, rad2}=2\pi\times \{840, 1030\}\,$kHz in the radial directions and $\omega_\text{ax}=2\pi\times 720\,$kHz in the axial direction. 
Multi-ion measurements are performed with a relaxed axial confinement, yielding centre-of-mass trapping frequencies of about $\omega_\text{COM,ax}=2\pi\times 220\,$kHz, and $\omega_\text{COM, rad1, rad2}=2\pi\times \{1080,1100\}\,$kHz, which allows for linear chains of up to 10 ions, extending over about \SI{60}{\micro m}. The same confinement was chosen regardless of the exact number of ions. The coupled motion of $N$ ions in the chain yields $3N$ motional modes. Low-frequency motional modes along the direction of clock interrogation cause loss of excitation probability due to thermal dephasing, leading to deteriorated clock stability~\cite{Keller2019}. For the axial motion, the COM mode has the lowest frequency. We perform 18\,ms of continuous resolved sideband cooling of this mode using the first and second red sidebands of the $-1/2\leftrightarrow -5/2$ transition before each clock interrogation. This additional step robustly restores the excitation probabilities to those obtained with the single-ion clock at high axial confinement. 

Each Zeeman component is interrogated with detunings $\pm \Delta\nu$ from its center frequency. Detunings of $\Delta \nu= 0.4/t$, with $t$ the interrogation time, approximately provide 50\% of the maximum excitation probability for the Fourier-limited Rabi lines. After completing a measurement cycle, the corresponding servo shifts the center frequency by $0.15\,(n_+-n_-)/N\times \Delta \nu $, with $n_+$($n_-$) the total number of successful excitations summed over all ions at positive (negative) detuning after two interrogations each, performed in the order [$-\Delta \nu$, $+\Delta \nu$, $+\Delta \nu$, $-\Delta \nu$], and $N$ the number of ions. The multi-ion clock is thus steered by the ensemble of all ions, and we verify that each ion in the chain contributes with equal overall excitation probability.

Position-dependent frequency shifts are obtained from the measured excitation imbalances $(n_+-n_-)_i$ and the known lineshape. We employ this technique to investigate two shift effects in an ion-site-resolved fashion:

First, magnetic field differences between the ions in the crystal lead to broadening of their collective spectrum. We apply a fixed compensation of a magnetic field gradient on the order of a few nT/mm along the axial direction using a dedicated set of magnetic field coils connected in anti-Helmholtz configuration. The residual gradient during clock operation is typically well below 0.1\,nT/mm. We do not observe nonlinear contributions or temporal variations during typical measurement runs spanning about two to four days.

\begin{figure*}
\centering
\includegraphics[width=0.75\textwidth]{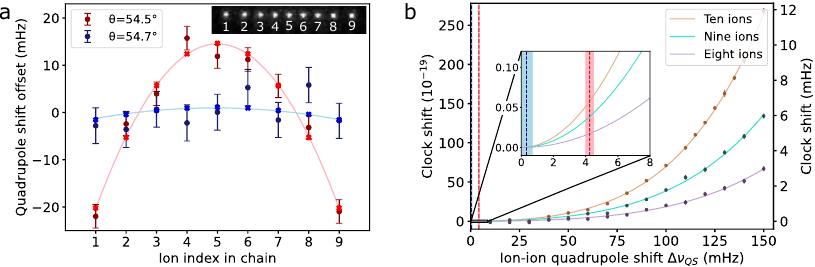}
\caption{Quadrupole shift variation along the ion chain. a) Ion-resolved quadrupole shift variation for a chain of nine ions. Red data points are results from a measurement with a misalignment of the angle $\theta$ between the magnetic field and the trap axis of about \SI{0.2}{\degree} from the optimal value of 54.7°. Blue data points are results from a typical measurement run with close to optimal magnetic field alignment. Error bars correspond to statistical uncertainties. Red and blue crosses show the result of a discrete fit of \autoref{eq:qs_var} to the data, yielding $\Delta\nu_{QS}=4.2(3)\,$mHz and $\Delta\nu_{QS}= 0.3(4)\,$mHz. Solid lines are quadratic fits to guide the eye. b) Simulated frequency shift due to line pulling as a function of the amount of quadrupole shift variation $\Delta\nu_{QS}$ across the ion chain. Each point corresponds to a Monte-Carlo simulation of the clock servo and the same interrogation times and weights as in the experiment. Error bars correspond to statistical uncertainties due to the finite number of simulated clock cycles. The solid lines are fits of the form $ax^2+bx^4$. The green and red dashed vertical lines and shaded areas correspond to the experimental examples in a). The latter corresponds to a clock shift of $0.038(5)\times 10^{-19}$.}\label{fig2}
\end{figure*}

Second, electric field gradients arise from the Coulomb-interactions between the ions. These are not homogeneous along the ion chain, and couple to the excited-state quadrupole moment $\Theta$. We can parametrize the resulting quadrupole shift of ion $i$ as~\cite{Keller2019}
\begin{equation}
    \begin{aligned}
  	\Delta\nu_{QS, i} =\Delta\nu_{QS}(\theta)  \, (m_J^2 -J(J+1)/3) \\ \times \left( \sum_{j\neq i} \frac{1}{\lvert{u_i-u_j}\rvert^3} \right)\label{eq:qs_var}
    \end{aligned}
\end{equation}
where the second term evaluates as -2/3 for $\lvert{m_J}\rvert=3/2$ and +10/3 for $\lvert{m_J}\rvert=5/2$, and $u_i$ denote the dimensionless equilibrium positions in the harmonic confining potential~\cite{James1998}. Here, we have defined 
\begin{equation}
    \Delta\nu_{QS}(\theta)=\frac{2m \omega_\text{ax}^2}{h\,e} \frac{3\cos^2\theta-1}{2} \frac{3}{J(2J-1)}\, \Theta
\end{equation}
where $\theta$ denotes the angle between the trap axis and the direction of the quantization axis, $h$ the Planck constant, and $e$ the elementary charge. This inhomogeneous ion-ion quadrupole shift results in asymmetric collective line shapes that cause a frequency offset of the multi-ion clock. Since $\Delta\nu_{QS}(\theta)$ becomes zero for $\theta=\arccos{1/\sqrt{3}}\approx$ \SI{54.7}{\degree}, the effect can be minimized by setting the magnetic field direction accordingly~\cite{Tan2019, Keller2019}. \autoref{fig2} shows the ion-resolved variation of the quadrupole shift for an intentional misalignment of the magnetic field direction by about \SI{0.2}{\degree} from the optimal condition, yielding $\Delta\nu_{QS}=4.2(3)\,$mHz, and for typical operating conditions, where we find $\lvert\Delta\nu_{QS}\rvert < 1\,$mHz. Monte Carlo simulations of the clock servo show that the resulting clock shift remains below $6\times10^{-21}$ for up to 10 ions. As the magnetic field angle is passively much more long-term stable than \SI{0.2}{\degree}, owing to a two-layer magnetic shielding, we do not apply a correction of this shift, but assign the largest possible shift of $0.1\times 10^{-19}$ as uncertainty .

In the following, we discuss the remaining frequency shifts and their contributions to the clock's uncertainty budget, which is summarized in \autoref{tab:table1}. Previous detailed discussions of systematic effects in \Sr clocks can be found in~\cite{Dube2013, Steinel2023, Lindvall2025}. By far the largest shift is caused by the blackbody radiation (BBR) of the room temperature environment. The effective temperature at the ion position is determined based on the reading of six resistive temperature sensors, of which two are soldered directly onto the ion-trap wafers. The sensors are calibrated at PTB with a k=2 uncertainty of 37\,mK. We use the model developed in~\cite{Nordmann2020} to characterize the influence of the temperature increase from the RF trap drive. We obtain an uncertainty of $50.7\,$mK on the effective temperature at the trap center for our operating conditions, corresponding to a fractional frequency uncertainty of $3.7\times10^{-19}$. 
The second contribution to the BBR shift uncertainty is knowledge of the transition's sensitivity to BBR. We use the recent measurement of the differential static polarizability in \cite{Lindvall2025a}, and its dynamic correction, as evaluated in~\cite{Dube2014}. Their uncertainties contribute a fractional uncertainty of $2.2\times 10^{-19}$, and $0.9\times 10^{-19}$, respectively, to the BBR shift. In total, we obtain a fractional uncertainty of $4.4\times10^{-19}$ of the BBR shift. 

To quantify shifts from thermal ion motion, we determine the temperature of a single ion after Doppler cooling with measurements of the relative excitation probability on the carrier and corresponding sideband transitions 
and obtain $T_\text{rad}=(T_\text{rad1}+T_\text{rad2})/2=0.91(22)\,$mK and $T_\text{avg}=(T_\text{rad1}+T_\text{rad2}+T_\text{ax})/3=0.80(16)\,$mK for the average radial and overall temperatures, with the uncertainties dominated by observed variations over the 10-month measurement period. 
We measured heating rates of $\dot{\bar{n}}_{\text{rad1, rad2, ax}}=\{0.3(1), 0.7(1), 3.6(3)\}$ quanta/s, which are not relevant on the timescales of the interrogation times. The combined second-order Doppler and Stark shift due to thermal motion is $-1.27\times 10^{-18}$ for a single ion, and the temperature uncertainty translates to a shift uncertainty of $2.6\times 10^{-19}$. All ions in a chain are homogeneously Doppler-cooled, with differences in fluorescence below 5\%. We assume that sideband cooling the axial COM mode to the motional ground state reduces the temperature of all $N$ ions by a factor of $(3N-1)/(3N)$, yielding a similar thermal shift of $-1.24(26)\times 10^{-18}$ for eight to ten ions.

Our trap drive frequency of $\Omega_{\text{RF}}=2\pi\times 14.242\,$MHz is close to the ``magic" RF frequency of about 14.39\,MHz~\cite{Lindvall2025a} where the quadratic scalar Stark shift and the time dilation shift from excess micromotion cancel~\cite{Dube2014}. Using the photoncorrelation technique~\cite{Keller2015b} we measure excess micromotion in all three directions and find a constant RF field amplitude of $65(10)\,$V/m in the axial direction. In the radial directions, micromotion compensation generally yields a RF field amplitude close to zero at the beginning of a measurement run. With measurement runs of up to several days and only intermittent monitoring of the micromotion, we use the maximal observed total radial amplitude after any measurement, about 107\,V/m, as the associated uncertainty. We obtain a combined frequency shift of $-0.3(0.9)\times10^{-19}$. By measuring excess micromotion of a single ion at different positions along the trap axis, and by by adding ions one by one and comparing the resulting collective photoncorrelation signals, we confirm that it is constant for different ion positions in the relevant range to within the stated uncertainty.

\begin{table}
\setlength\tabcolsep{-1pt}
\sisetup{table-alignment-mode = marker}
\centering
\renewcommand{\arraystretch}{1.2}
\begin{tabular}{@{}l S S@{}}

 {Effect}&{shift}&{uncertainty} \\
 \B&{$\left(10^{-19}\right)$}&{$\left(10^{-19}\right)$}\\ \hline
 Blackbody radiation\footnotemark[1]&5382.2&4.4  \\
 Thermal motion&-12.4&2.6\\
 Excess micromotion&-0.3&0.9\\
 Collisions&0&0.9\\
 Quadrupole shift inhomogeneity &0&0.1\\
 Servo error (Residual 1st order Zeeman)& 0& 0.1\\
 2nd-order Zeeman (static B-field)\footnotemark[2] &0.6&<0.1\\
 2nd-order Zeeman (BBR) &-0.1&<0.1\\
 2nd-order Zeeman (trap drive) &2.7&<0.1\\
 Light shift (674 nm) &0.2&<0.1\\
 Ellipticity-induced light shift (674 nm) &-0.7&<0.1\\\hline
 Total&5372.2 &5.3\T \\
\end{tabular}
\caption{\label{tab:table1}Frequency shifts and corresponding uncertainties for the \Sr clock operating with eight to ten ions. Single-ion operation yields a higher systematic uncertainty of $6.5\times10^{-19}$ due to a $3.9\times10^{-19}$ uncertainty of the collisional shift, with otherwise equal uncertainties.}
\footnotetext[1]{Time-resolved correction based on resistive temperature sensors. Value given is for a typical temperature of 296.75\,K.}
\footnotetext[2]{Time-resolved correction based on resonant frequencies and known Zeeman sensitivities. Value given is for a typical magnetic field strength of 3.0\,\textmu T.}
\end{table}

\noindent

Collisions of the ions with background gas atoms or molecules may lead to frequency shifts~\cite{Hankin2019}.
We measure a background-gas collision rate of $1.16(7)\times10^{-3}$/s per ion by observing reordering events in a mixed-species ion crystal containing a different, non-fluorescing Sr\textsuperscript{+} isotope. While almost all collisions remain undetected when operating with a single ion, as indicated by the absence of invalid clock cycles, more than 78\% of collisions lead to crystal melting followed by invalid clock cycles when operating with five or more ions. 
Following the simple treatment in~\cite{Rosenband2008}, and only including undetected collisions, yields a fractional frequency uncertainty of $3.9\times10^{-19}$ in the single-ion case and $0.9\times10^{-19}$ for the multi-ion clock. 
Thus operation with multiple ions is not just beneficial for a clock's instability, but also its systematics.

The 674\,nm clock laser is pre-stabilized by a routinely operated single-ion \Yb clock at a frequency comb, which is running with an instability of the same order as the \Sr clock (see Methods section for details). This makes servo errors from a drift of the local oscillator negligible during operation of the \Yb clock. The only remaining servo error comes in the form of a residual first-order Zeeman shift due to magnetic field variations. We evaluate this contribution to be below $0.1\times10^{-19}$.

Further details, and a discussion of the remaining shifts -- with systematic uncertainties at or below the $0.1\times 10^{-19}$-level -- can be found in the Methods section. 

\begin{figure*}
\centering
\includegraphics[width=0.75\textwidth]{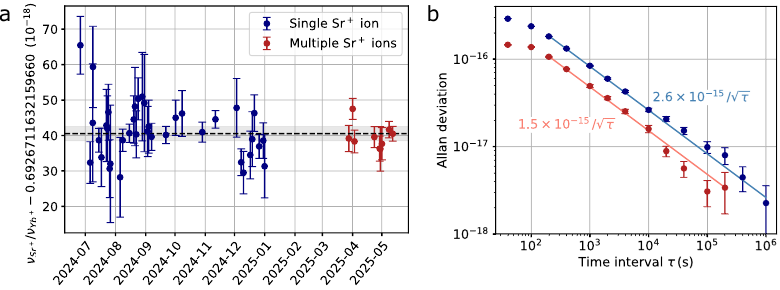}
\caption{a)  Measurements of the ratio between the frequencies $\nu_{\text{Sr}^+}$ of the $^2S_{1/2}\leftrightarrow{}^2D_{5/2}$ transition in \Sr and $\nu_{\text{Yb}^+}$ of the ${}^2S_{1/2} (F=0)\leftrightarrow {}^2F_{7/2} (F=3)$ transition in \Yb. Data points shown in blue were obtained with a single \Sr ion, while data points shown in red correspond to measurements with eight to ten \Sr ions. Error bars correspond to the statistical uncertainties of the individual measurements, assuming white frequency noise for each entire measurement interval. We find $\chi^2_\text{red}\approx1.1$ and an overall statistical uncertainty of $0.9\times10^{-18}$. The dashed line corresponds to the average of all measurements, weighted by their statistical uncertainties, and the grey shaded area depicts the combined overall uncertainty of $2.9\times 10^{-18}$. c) Measurement instability as characterized by the Allan deviation for operation of the \Sr clock with a single ion (shown in blue) and eight to ten ions (shown in red). The error bars are the statistical uncertainties. The solid lines depict fits according to the $1/\sqrt{\tau}$ scaling for white frequency noise. We evaluated about $4.55\times10^6$\,s of single-ion data and $1.22\times10^6$\,s of multi-ion data, reaching statistical uncertainties of $1.2\times10^{-18}$ and $1.4\times10^{-18}$ respectively.}\label{fig3}
\end{figure*}

We compare the new optical clock based on \Sr to an established single-ion clock based on the ${}^2S_{1/2} (F=0)\leftrightarrow {}^2F_{7/2} (F=3)$ electric octupole (E3) transition in \Yb~\cite{Huntemann2016}. This clock has previously been evaluated with a systematic uncertainty of $2.7\times 10^{-18}$~\cite{Sanner2019} and has demonstrated an instability of $1.0\times 10^{-15}/\sqrt{\tau (\text{s})}$ for an averaging time $\tau$~\cite{Sanner2019, Dorscher2021}. For completeness, the full uncertainty budget of the \Yb clock under recent operating conditions is provided in the methods section. The frequency ratio data is shown in \autoref{fig3}. The statistical uncertainty from all data is $9.1\times 10^{-19}$, and we find $\chi_\text{red}^2\approx 1.1$, compatible with white frequency noise. This demonstrates a long-term stability of the \Yb clock well below its systematic uncertainty. The frequency ratio $\nu_{\text{Sr}^+}/\nu_{\text{Yb}^+}$ of the two transitions is 0.6926711632159660405(20), where a gravitational redshift of $-2.620(31)\times10^{-17}$ between the two clocks has been taken into account. We obtain a combined uncertainty of $2.9\times10^{-18}$, limited by the single-ion clock. This constitutes a significant step towards a future optical definition of the SI unit second, as it is one of very few optical frequency ratios between different atomic transitions that fulfill the criterion of an overall fractional uncertainty $\le 5\times10^{-18}$~\cite{Dimarcq2024}. The first one, with an overall uncertainty of $4.4\times10^{-18}$, was that measured between the same \Yb single-ion clock, and the \textsuperscript{115}In\textsuperscript{+} Coulomb crystal clock at PTB~\cite{Hausser2025}. 

Our result is consistent at the $1\sigma$ level with that obtained in~\cite{Steinel2023}, when accounting for the different value of the differential polarizability that was used in the correction of the BBR shift, and reduces the uncertainty of $\nu_{\text{Sr}^+}/\nu_{\text{Yb}^+}$ by almost an order of magnitude. 
With a single \Sr ion, we find a measurement instability of $2.6\times10^{-15}/\sqrt{\tau (s)}$, indicating an instability of $2.4\times10^{-15}/\sqrt{\tau (s)}$ of the \Sr clock. For $N$ ions, one would expect a reduction of the clock's instability by a factor of $1/\sqrt{N}$. We find that the dead time associated with frequent re-cooling after crystal melting limits the instability of the \Sr clock operating with eight to ten ions to about $1.1\times10^{-15}/\sqrt{\tau (s)}$. Currently, re-cooling of the crystal takes about 4\,s, and the instability could be improved further by making this process more efficient. Lower collision rates, for example in a cryogenic environment, would also contribute to an improved instability. 

To summarize, we have reported on an optical clock based on up to ten \Sr ions with a systematic uncertainty of $5.3\times10^{-19}$. We show that the effects of varying external perturbations along the extended ion chain are suppressed to below the $10^{-20}$ level. Notably, operation with multiple ions compared to a single ion does not just reduce the instability of the clock, but also its systematic uncertainty because of the improved detection of collisions. We have compared the \Sr clock to a single-ion clock based on the electric octupole transition in \Yb, obtaining a combined fractional uncertainty of $2.9\times10^{-18}$ on the resulting frequency ratio, fulfilling a criterion towards a future optical definition of the second. 
This work establishes multi-ion clocks as systems that can achieve exceptionally low systematic uncertainties, and instabilities beyond the limitations of a single ion.  
 The developed techniques can also be applied to other ion species with suitable clock transitions. 
\newpage

\begin{acknowledgments}
We thank Uwe Sterr, Thomas Legero, and Jialiang Yu for providing an ultrastable reference oscillator. We are grateful to Christian Sanner, Thomas Lindvall, Jonas Keller, and Piet Schmidt for helpful discussions. We thank Martin Menzel, Nele Wendt, and Burghard Lipphardt for experimental support. We thank the PTB experimental instrumentation department for manufacturing of the ion trap components. We thank William Eckner for helpful comments on the manuscript.
This work was supported by the Max Planck-RIKEN-PTB Center for Time, Constants and Fundamental Symmetries, by the Deutsche Forschungsgemeinschaft (DFG, German Research Foundation) under SFB 1227 DQ-mat–Project-ID 274200144–within projects B02 and B03, and under Germany’s Excellence Strategy – EXC-2123 QuantumFrontiers – 390837967, and by the project 23FUN03 HIOC. The project (23FUN03 HIOC) has received funding from the European Partnership on Metrology, co-financed from the European Union’s Horizon Europe Research and Innovation Programme and by the Participating States. 
\end{acknowledgments}

\section*{Author contributions}

M.F., M.R.S., J.J. and N.H. built the experimental apparatus. T.E.M. provided the ion trap design and D.B. contributed to its construction. M.R.S. and M.F. developed the experimental control software and performed shift characterization measurements. M.F., M.R.S., and N.H. operated the optical clocks and analyzed the frequency ratio data. M.F. prepared the manuscript with input from all authors. N.H. and E.P. conceived the project and supervised the work. 

\section*{Methods}
\subsection{Measurement infrastructure for optical frequency comparison}

\begin{figure*}
	\centering
	\includegraphics[width=0.7\linewidth]{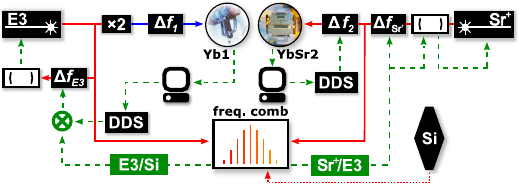}
	\caption{Key components in the measurement of $\nu_{\text{Sr}^+}/\nu_{\textrm{E3}}$. Yb1 refers to the \Yb single-ion clock, while YbSr2 refers to the apparatus realizing the new \Sr multi-ion clock. Optical paths are depicted by red and blue lines; electric signals by dashed green lines. The E3 and Sr\textsuperscript{+} clock lasers are stabilized to optical cavities, and the E3 laser light is frequency doubled before being sent to the ion trap. The short-term stability of an ultrastable silicon cavity (Si)~\cite{Matei2017} is transferred via a frequency comb to the E3 clock laser~\cite{Benkler2019}. The Sr\textsuperscript{+} laser frequency is controlled to provide a constant frequency ratio relative to the E3 laser at a frequency comb via the frequency offset $\Delta f_{\text{Sr}^+}$ and the cavity length, thereby inheriting the Si short term stability from the E3 laser. For simplicity, only one frequency comb is depicted, although two coherently linked frequency combs are employed. Spectroscopy of the E3 clock transition steers the corresponding laser frequency via direct digital synthesis (DDS) providing the frequency offset $\Delta f_\text{E3}$, while the frequency of the Sr\textsuperscript{+} laser is corrected by an additional offset to $\Delta f_2$ directly before the ion trap.}
	\label{fig:YbvsSr_setup}
\end{figure*}

Key components involved in the optical frequency comparison are depicted in \autoref{fig:YbvsSr_setup}. The Sr\textsuperscript{+} probe laser frequency is stabilized with a fixed ratio $R_0$ to that of the \Yb E3 probe laser at a frequency comb. The servo running on the E3 transition steers the probe laser directly, while the servo of the Sr\textsuperscript{+} clock steers a local frequency offset $\Delta f_2$ via an AOM close to the ion trap, which is recorded and can be used to calculate the frequency ratio via
\begin{equation}
	\frac{\nu_\text{Sr\textsuperscript{+}}}{\nu_\text{E3}}= R_0\left(1 -\frac{\Delta f_\text{2}}{\nu_\text{Sr\textsuperscript{+}}} - \frac{\Delta \nu_\text{Sr\textsuperscript{+}}}{\nu_\text{Sr\textsuperscript{+}}} + \frac{\Delta \nu_\text{E3}}{\nu_\text{E3}}\right)\, , \label{eq:sre3ratio}
\end{equation}
where the frequency offsets $\Delta \nu_\text{E3}$ and $\Delta \nu_\text{Sr\textsuperscript{+}}$ can be obtained from the respective uncertainty budgets. For the \Sr clock, time-resolved corrections for the BBR shift and second-order Zeeman shift are applied, so that $\Delta \nu_\text{Sr\textsuperscript{+}}$ and $\Delta f_2$ are both time-dependent. 

\subsection{Uncertainty budget of the single-ion \textsuperscript{171}Yb\textsuperscript{+} clock}

For the measurements presented here, the operating parameters of the single-ion \Yb clock Yb1 result in the uncertainty budget shown in~\autoref{tab:table2}. 

\begin{table}[h!]
\sisetup{table-alignment-mode = marker}
\centering
\renewcommand{\arraystretch}{1.2}
\begin{tabular}{@{}l S S@{}}

 {Effect}&{shift}&{uncertainty} \\
 \B&{$\left(10^{-18}\right)$}&{$\left(10^{-18}\right)$}\\ \hline
Blackbody radiation &  -71.6 & 1.8 \\
Second-order Doppler  & -2.3 & 1.5 \\
Light shift (probe)  &  0 & 0.8 \\
Trap-induced AC Zeeman & 0.6 & 0.6 \\
Trap RF Stark  &-0.8 & 0.6 \\
Electric quadrupole &  -4.7 & 0.5 \\
Background gas collisions &  0 & 0.5 \\
Quadratic Zeeman (DC) &  -28.8 & 0.2\\
Servo error& 0 & 0.2 \\
Phase chirp  & 0 & 0.1 \\ \hline

 Total&-107.6 & 2.7\T \\
\end{tabular}

\caption{\label{tab:table2}Frequency shifts and corresponding un\-cer\-tain\-ties for the \Yb single-ion clock opera\-ting on the ${}^2S_{1/2} (F=0)\leftrightarrow {}^2F_{7/2} (F=3)$ electric octupole transition.}
\end{table}

Systematic uncertainties remain unchanged from the previous evaluation reported in~\cite{Sanner2019}. We note that compared to the shift corrections applied in recently reported optical frequency ratios involving this apparatus~\cite{Hausser2025, Lindvall2025b}, the value of the overall applied frequency correction is smaller by $0.8\times10^{-18}$. The differences are a slightly higher temperature of the environment, leading to a larger BBR shift, a slightly smaller measured quadrupole shift, and the newly implemented correction of the AC Zeeman shift from the trap drive.

\subsection{Thermal environment and temperature uncertainty}

To characterize the thermal environment of the ions, we employ six resistive temperature sensors, which have been calibrated with a $k=2$ uncertainty of $37\,$mK at PTB, corresponding to a  $1\sigma$-uncertainty of $\sigma_{T_\text{sensor}}=18.5\,$mK. Two of the sensors are soldered directly onto the RF-carrying inner trap wafers, while the remaining four are distributed on the outside of the vacuum chamber. A linear fit to each sensor's calibration points is used to interpolate the temperature readings. We pre-selected sensors for calibration that showed no relevant variation of the measured temperatures after heating and cooling processes, which is important since the calibration was performed before soldering the sensors to the trap chips. Low-pass filtering is employed to mitigate the effects of RF-pickoff on the temperature readings.

When applying a radiofrequency voltage to the trap, the temperatures of the trap increases with respect to the background temperature of the chamber $T_{\text{ch}}$. Additionally, the temperature sensor on the outside of the RF feedthrough also shows an elevated temperature, leading to increased inhomogeneity between the four temperature sensors on the outside of the chamber. When no RF power is supplied to the ion trap and the system is in thermal equilibrium, the readings from all sensors agree to within less than 35\,mK. 

During clock operation, the temperature rise of the trap electrodes due to the applied RF power has to be considered, since they make up a significant fraction of the solid angle around the ions. The ion trap we employ has been specifically designed to provide a low temperature rise under RF, and extensive thermal modeling for this trap geometry has been developed~\cite{Nordmann2020}.

The effective temperature experienced by the ions 
\begin{equation}
	T_{\text{ions}}=T_{\text{ch}}+\Delta T_{\text{ions}}
\end{equation}
is extracted from the temperature of the chamber $T_{\text{ch}}$, which we obtain as the mean of the four temperature sensors on the vacuum chamber, and the effective temperature increase $\Delta T_{\text{ions}}$ experienced by the ions inside the trap, which can be obtained from the average~\cite{Nordmann2020}:
\begin{equation}
	\Delta T_{\text{ions}}=\frac{\alpha\,\Delta T_1+\beta\,\Delta T_2}{2}\,,
\end{equation}
with $\alpha=0.43(10)$, $\beta=0.32(10)$, and $\Delta T_{1,2}= T_{1,2}-T_{\text{ch}}$ the temperature differences between $T_1$ and $T_2$ read out from the sensors on the trap wafers and the vacuum chamber temperature. Here, $T_1$ denotes the sensor closer to the trap's carrier board, and $T_2$ that further away from it.

The overall resulting uncertainty of the temperature experienced by the ions is then
\begin{equation}\label{eq:temp_unc}	
	\begin{aligned}
		\sigma_{T_\text{ions}}= &\left[ \left(\frac{1}{2}\Delta T_1\,\sigma_\alpha\right)^2\right.  + \left(\frac{1}{2}\Delta T_2\,\sigma_\beta\right)^2\\
        &+ \left(\left(1-\frac{\alpha}{2}-\frac{\beta}{2}\right)\sigma_{T_\text{ch}}\right)^2 \\
		&\left. +\left(\frac{1}{2}\alpha \,\sigma_{T_1}\right)^2+ \left(\frac{1}{2}\beta\, \sigma_{T_2}\right)^2\,\right] ^{1/2}\,.
	\end{aligned}	
\end{equation}

The uncertainty of the chamber temperature is given by
\begin{equation}
	\sigma_{T_\text{ch}}=\sqrt{\sigma_{T_{\text{ch,stat}}}^2+ \sigma_{T_{\text{sensor}}}^2}
\end{equation}
where the uncertainty due to its inhomogeneity is
\begin{equation}
	\sigma_{T_{\text{ch,stat}}}= \left(\text{max}(\{T_i\})-\text{min}(\{T_i\})\right)/\sqrt{12}\, ,
\end{equation} 
with $\{T_i\} = \{T_3,...,T_6\}$.

For the work presented here, with radial secular frequencies around 1\,MHz for a single \Sr ion, we find $\left(\text{max}(\{T_i\})-\text{min}(\{T_i\})\right)\le 150\,\text{mK}$ and therefore $\sigma_{ T_{\text{ch}}}= \SI{47}{mK}$. Using the observed upper bounds of $\Delta T_1\approx 570\,$mK and $\Delta T_2\approx 590\,$mK yields an overall temperature uncertainty of $\sigma_{ T_{\text{ions}}}= 50.7\,$mK. Here, the first three terms in \autoref{eq:temp_unc} dominate, each contributing close to 1/3 of the uncertainty. The temperature uncertainty could be drastically reduced by actively cooling the RF feedthrough and/or the connected ion trap, thereby reducing the RF-induced temperature increases $\Delta T_{1,2}$, as well as the inhomogeneity of the chamber temperature, which is currently dominated by the temperature increase around the RF feedthrough.

 All temperature sensors are automatically read out every second, and the results are logged in a database. They are used to calculate a time-resolved correction of the BBR shift with a temporal resolution of 10\,s.

\subsection{Excess micromotion}

The fractional net frequency shift from excess micromotion to first order in the trap drive frequency is given by~\cite{Dube2013, Dube2014}
\begin{equation}
	\frac{\Delta \nu_\text{EMM}}{\nu_0}=-\frac{1}{2} \left [  \frac{\Delta \alpha_0}{h \nu_0} + \left ( \frac{e}{m\Omega_\text{RF} c} \right )^2 \right ] \langle E^2 \rangle \,
\end{equation}
with $\nu_0$ the transition frequency, $\Delta \alpha_0$ the differential scalar polarizability, $m$ the mass of a \Sr ion, $c$ the speed of light, and $\langle E^2 \rangle$ the mean square electric field from the trap drive experienced by the ions. Here, the first term is the scalar Stark shift, and the second term the second-order Doppler shift. Our trap drive frequency of $\Omega_\text{RF}=14.242\,$MHz is close to the ``magic" RF frequency of about 14.39\,MHz~\cite{Lindvall2025a}, where these two shifts cancel completely. Our experimental parameters correspond to a suppression of each of the two shifts via their mutual cancellation by about a factor of 53. Contributions from higher-order trap drive harmonics are neglected since they are suppressed by almost three orders of magnitude for our trapping parameters. The tensor Stark shift is canceled by the averaging scheme.

\subsection{Servo error}

Typically, the dominant servo error results from drift of the clock laser frequency. As the Sr\textsuperscript{+} clock laser frequency is stabilized to that of the E3 probe laser, it is drift-free during operation of the \Yb single-ion clock steering the E3 laser frequency. During intervals without steering from the \Yb clock, the small drift from the ultrastable silicon cavity at PTB~\cite{Matei2017} becomes relevant. The corresponding servo error can be mitigated by implementing drift correction into the servo sequence of the \Sr clock, as is currently the case for the \Yb clock. Because of the large operational uptime of the \Yb single-ion clock, this has not been necessary yet.

As we do not observe any variation of the quadrupole shift during measurement runs, magnetic field variations are the leading remaining possible cause of servo errors. Drifts of the magnetic field strength would lead to an imperfect cancellation of the first order Zeeman shift in the center frequency resulting from all servos. The effect is suppressed at multiple levels: Firstly, each Zeeman component is interrogated in a $[-\Delta \nu,+\Delta \nu,+\Delta \nu,-\Delta \nu]$ pattern~\cite{Lindvall2023}. Secondly, the order of interrogating the different Zeeman components during one clock cycle yields a further non-trivial suppression. 

We simulate the effect a linear magnetic field drift has on our specific servo sequence, including the employed interrogation times, order of interrogation, as well as the dead time, and find a fractional frequency offset of about $1.0\times 10^{-5}/[(\Delta B/\Delta t)(\text{T/s})]$, depending on the drift rate $\Delta B/\Delta t$. To estimate the overall effect of the observed noise, we numerically calculate the first derivative of the observed magnetic field strength on 200\,s time steps and integrate the accumulated error over the time steps for a selection of typical measurement runs. We obtain fractional frequency shifts below $0.1\times 10^{-19}$. Additional offsets due to nonlinear magnetic field changes on these timescales are expected to be even smaller. 

As we do not switch the magnitude or direction of the applied static field during the clock sequence, and also do not employ any mechanically changing components in the vicinity of the ions, we do not see a possible source of cycle-synchronous magnetic field changes.

\subsection{Second-order Zeeman shifts}
At our current operating conditions, the oscillating magnetic field from the trap drive contributes the largest second-order Zeeman shift. Here, there is no dependence of this shift on the direction of the quantization axis due to our averaging scheme canceling the tensor term. Following the approach developed in~\cite{Gan2018}, we find the oscillating magnetic field amplitude of 6.252(54)\,\textmu T. With an average second-order Zeeman sensitivity of 3.122\,\textmu Hz/\textmu $\text{T}^2$~\cite{Dube2013}, we obtain a frequency shift of $2.74(5)\times10^{-19}$ from the oscillating magnetic field.

The static magnetic field of about 3\,\textmu T used to define the quantization axis is determined in a time-resolved fashion from the servos running on the individual Zeeman components and their respective first-order Zeeman sensitivities. The corresponding second-order Zeeman shift is then also calculated and applied in a time-resolved fashion, with an average shift of $0.63 \times 10^{-19}$ and a $10^{-22}$ uncertainty.

Additionally, the magnetic field from the blackbody radiation of the environment has to be considered. The associated second-order Zeeman shift is influenced by the coupling of the radiation to the ${}^2D_{5/2}\leftrightarrow {}^2D_{3/2}$ fine-structure transition, and has been calculated to be $-0.11\times10^{-19}$ at room temperature in~\cite{Tang2024}.  

\subsection{Light shifts from clock laser}
All laser beams except for the 674\,nm clock beam are blocked by mechanical shutters during clock interrogation. Therefore, we only need to consider the light shift from the 674~nm clock laser itself. To assess the shift on the $^2S_{1/2}\rightarrow {}^2D_{5/2}$ clock transition frequency from off-resonant coupling to electric-dipole allowed transitions, related to the differential scalar and tensorial polarizabilities~\cite{Jiang2009}. 
The  tensorial contribution does not cancel completely, as we interrogate the different Zeeman components with different intensities.

We use the relevant differential polarizabilities with an estimated uncertainty of $25\%$ from~\cite{Madej2004}, yielding a convenient expression for the light shift from the interrogation laser in terms of the intensity $I_0$ at the position of the ion(s):
\begin{widetext}
\begin{equation}
	\Delta \nu_{674\,\text{nm}} = I_0 \left( 0.525 \\
	+ 0.064\, (3 \cos^2\beta-1)\,\frac{3m_J^2-J(J+1)}{J(2J+1)}\right)\,\text{mHz}/(\text{W}/\text{m}^2)\,.
\end{equation}
\end{widetext}

Here, $m_J$ and $J$ refer to the excited-state quantum numbers. The intensity is obtained by measuring the laser power before and after the vacuum chamber for different intermediate powers and extrapolating to the (very small) powers used during the clock sequence. From measuring the beam diameter at two positions outside the vacuum chamber, we extract an estimated beam waist radius at the ion position of $100(20)$\,\textmu m. We obtain intensities of $I_0(\pm5/2)\approx 0.031\, \text{W}/\text{m}^2$ and $I_0(\pm3/2)\approx 0.011\, \text{W}/\text{m}^2$ during the interrogation pulses for the $\pm 1/2\rightarrow\pm 5/2$ and $\pm 1/2\rightarrow\pm 3/2$ Zeeman pairs, respectively. 

In our configuration, the clock beam is aligned along the trap axis and horizontally polarized, so that typically $\beta\approx \SI{90}{\degree}-\SI{54.7}{\degree}= \SI{35.3}{\degree}$, and for some of the single-ion measurements we used $\beta\approx\SI{90}{\degree}-\SI{61.4}{\degree}= \SI{28.6}{\degree}$. We calculate the total shift for each of the two relevant Zeeman pairs, and perform a weighted average according to our determination of the center frequency (weight 5/6 for $\pm 1/2\rightarrow\pm 3/2$ and 1/6 for $\pm 1/2\rightarrow\pm 5/2$). We obtain a fractional frequency shift of about $0.17\times10^{-19}$ for $\beta= \SI{35.3}{\degree}$ and $0.18\times10^{-19}$ for $\beta= \SI{28.6}{\degree}$, each with an estimated uncertainty of 30\%, meaning the difference between the two cases is negligible.

Additionally, each Zeeman component of the clock transition is shifted by off-resonant coupling to all other Zeeman components that it shares an energy level with. For a linearly polarized probe laser beam, these shifts cancel in the average of a Zeeman pair. Residual ellipticity in the polarization, however, leads to imbalanced coupling strengths, resulting in a remaining shift after averaging~\cite{Yudin2023, Lindvall2025}. We find an imbalance in the Rabi frequencies of the Zeeman pairs of about 15\% for $\Delta m=\pm1$ and 17\% for $\Delta m=\pm2$. We use the measured Rabi frequencies directly to calculate the ellipticity-induced shift for our experimental parameters, summing over all relevant terms for each Zeeman component. We obtain a shift of $-0.66\times10^{-19}$ for a magnetic field strength of \SI{3}{\micro T}, with an estimated uncertainty of about 10\%. We note that this shift can be avoided in future measurements by carefully minimizing residual ellipticity via modifying the polarization to balance the Rabi frequencies of the two components in a Zeeman pair. \\
\vspace{1.1 mm}
\subsection{AOM phase chirp}
The acousto-optic modulator (AOM) used for clock interrogation is repeatedly switched on and off during the experimental sequence. This leads to temperature changes in the AOM crystal, which in turn cause optical path length changes~\cite{Kazda2016}. We actively stabilize the optical path length between the clock laser system and a point about 15\,cm from the vacuum chamber based on interferometry with the retroreflected 0th diffraction order of the AOM used for clock interrogation. The bandwidth of this stabilization is about 10\,kHz, and consequently any AOM phase chirp that is common mode to both diffraction orders will be suppressed by a factor of about 1000 for a pulse of 100\,ms. 

At full power, the observed phase error due to AOM chirp in \cite{Kazda2016} is 9\,mrad for an ``off" time of 2\,ms. This means that for typical ``off" times of 40\,ms in our experiment, and with a similar duty cycle, we expect 180\,mrad at full power, corresponding to a fractional frequency offset of $6\times10^{-16}$. We use the AOM at more than 40\,dB below full diffraction efficiency, corresponding to a factor 10\,000 in power. Linear scaling yields an estimated fractional shift below $6\times10^{-20}$, even without considering the additional suppression from the fiber length stabilization.

\bibliography{manuscript.bib}

\end{document}